\documentstyle[11pt,newpasp,psfig,twoside]{article}
\def\edcomment#1{\iffalse\marginpar{\raggedright\sl#1\/}\else\relax\fi}
\reversemarginpar

\begin{document}
\title{VLT spectra of the companion candidate Cha H$\alpha$ 5/cc 1}
\author{Ralph Neuh\"auser}
\affil{MPE, D-85740 Garching, Germany}
\author{Eike Guenther}
\affil{TLS, D-07778 Tautenburg, Germany}
\author{Wolfgang Brandner}
\affil{ESO, D-85748 Garching, Germany}

\begin{abstract}
We obtained optical and infrared spectra of Cha H$\alpha$ 5/cc 1, 
a faint possibly sub-stellar companion candidate
next to the M6-type brown dwarf candidate Cha H$\alpha$ 1 in Cha I,
using FORS1 and ISAAC at the VLT. The VRIJHK colors of
Cha H$\alpha$ 5/cc 1 are consistent with either an L-type companion
or a K-type background giant. Our spectra show that the companion candidate
actually is a background star.
\end{abstract}

\section{Introduction}

Twelve M6- to M8-type objects called Cha H$\alpha$ 1 to 12 were found in
deep infrared, H$\alpha$, and X-ray surveys (Comer\'on et al. 2000). 
To search for faint visual companions around these
bona-fide and candidate brown dwarfs, we have taken deep images
with HST WFPC2 (R, I, H$\alpha$), VLT FORS1 (VRI), and NTT SofI (JHK$_{\rm s}$)
and detected one particulary promising companion candidate:
Cha H$\alpha$ 5/cc 1, a companion candidate 1.5 arc sec off Cha H$\alpha$ 1,
is 3.8 to 4.7 mag fainter than the primary and
its colors are consistent with an early- to mid-L spectral type.
Assuming the same distance, absorption, and age as for the primary, 
the faint object would have a mass of 3 to 15 Jupiters according
to Burrows et al. (1997) and Chabrier \& Baraffe (2000) models.
The probability for this companion candidate to be an unrelated fore- or background object
is $\le 0.7\%$, its VRIJHK colors are marginally consistent with a strongly reddened
background K giant. These results are published in Neuh\"auser et al. (2002).

Even with the best currently achievable astrometric precision (few mas),
we would have to wait several years for checking whether this visual pair is a
common proper motion pair. Alternativelly, and faster, one can check by
spectroscopy whether the faint companion candidate is either
an L-type companion or a reddened K-type background giant.

\section{Observations}

Spectra were taken with ISAAC and FORS1 at the VLT in early 2002.

The H- and K-band spectra show that the companion candidate is as red
as concluded before from its VRIJHK colors.
However, in the K-band spectrum of the companion candidate, the sharp drop of
the quasi-continuum at $\ge 2.3~\mu m$ due to CO lines is not present.
This drop and the CO lines can be seen in the spectrum of Cha H$\alpha$ 5,
but should be even stronger in an L-type object.

In the optical spectrum, the three CaII lines at 8661.7, 8541.7,
and 8497.6 \AA , which are typical for K stars, but are not present in L-dwarfs, are
detected. Also, the typical features of L-dwarfs, like K, Rb, Cs, FeH, and CrH lines
are not detected. In Fig. 1, we compare the companion candidate's spectrum with a
typical K-type giant star showing the similarities in absorption lines and their 
relative strength (companion candidate not corrected for strong reddening).

\begin{figure}
\vbox{\psfig{figure=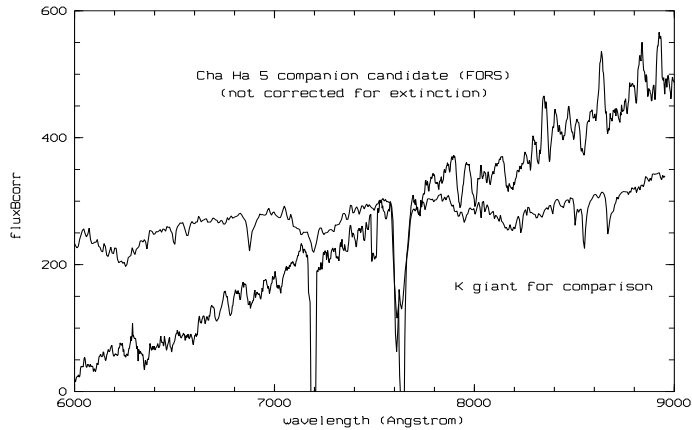,width=10cm,height=6cm,angle=270}}
\caption{VLT/FORS spectrum (not corrected for reddening) of Cha H$\alpha$/cc 1 
compared to a K-type giant.}
\end{figure}

From the previous imaging data, we concluded that the companion candidate is 
either an unabsorbed L-type companion or a reddened K-type background giant,
with the former alternative being much more likely ($99.3 \%$).
From the follow-up spectra, we conclude that this candidate is
a background star, probably a K-type giant.

\end{document}